\documentclass[showpacs,nofootinbib,showkeys,eqsecnum,prd,aps,preprint]{revtex4-1}

\usepackage{amssymb}
\usepackage{amsmath}
\usepackage{amsfonts}
\usepackage{graphicx}
\usepackage{bm}
\usepackage[T1]{fontenc}
\usepackage{graphicx}
\usepackage{hyperref}

\begin{document}

\title{Interaction between two point-like charges in nonlinear electrostatics}

\author{A.~I.~Breev${}^{b,c}$}
\email{breev@mail.tsu.ru}

\author{A.~E.~Shabad${}^{a,b}$}
\email{shabad@lpi.ru}

\date{\today }

\begin{abstract}
We consider two point-like charges in electrostatic interaction between them
within the framework of a nonlinear model, associated with QED, that
provides finiteness of their field energy. We find the common field of the two charges in a dipole-like approximation, where the separation between them $R$ is much smaller than the observation distance $r:$ with the linear accuracy with respect to the ratio $R/r$, and
in the opposite approximation, where $R\gg r,$ up to the the term quadratic in the ratio $r/R$. The consideration fulfilled proposes the law $a+b R^{1/3}$ for the energy, when the charges are close to one another, $R\rightarrow 0$. This leads to the singularity of the force between them to be $R^{-2/3}$, which is weaker than Coulomb law $R^{-2}$.
\end{abstract}

\affiliation{
$^{a}$P. N. Lebedev Physical Institute, 53 Leninskiy prospect, 119991,
Moscow, Russia\\
$^{b}$Department of Physics, Tomsk State University, Lenin Prospekt 36, 634050, Tomsk, Russia\\
${}^{c}$Department of Higher Mathematics and Mathematical Physics, \\
Tomsk Polytechnic University, Lenin Prospekt, 30, 634034, Tomsk, Russia
}

\maketitle

\section{Introduction}
\label{s:intro}
Recently a class of nonlinear electrodynamic models was studied \cite%
{CosGitSha2013a} wherein the electrostatic field of a point charge is, as
usual, infinite in the point, where the charge is located, but this
singularity is weaker than that of the Coulomb field, so that the space
integral for the energy stored in the field converges (and also the scalar
potential is finite in the position of the charge \cite{AdoGitShaShi2016}).
This class unites Lagrangians \cite{AdoGitShaShi2016} - \cite{Ependiev} that
grow with the field invariant $\mathfrak{F=}\left( B^{2}-E^{2}\right) /2$
faster than $(-\mathfrak{F)}^{w},w>$ $3/2$ (see Ref. \cite%
{AdoGitShaShi2016} also for a subtler estimate of the boundary of
the necessary growth). Thus, the simple quadratic effective
Lagrangian first considered in a different aspect in
\cite{Kruglov2008} is also included into the class under
consideration.

The infiniteness of the field near the charge distinguishes the class under
consideration from many other models (see the papers \cite{Kruglov} and
references therein) with finite self-energy of the point charge, allied to
their famous prototype, the Born-Infeld model \cite{Born-Infeld}, where the
finiteness of the field is achieved at the cost of square-root
nonanalyticity of the Lagrangian that supplies an infinity to the Maxwell
equation. The most popular application \cite{Kruglov}, \cite{Hendi} of these
models is to combine them with the General Relativity in order to study
their effect on the initial singularity and on the evolution of the
Universe. In contrast to the Born-Infeld model, the models from the class of
Ref. \cite{CosGitSha2013a} refer to nonsingular Lagrangians that follow for
instance from the Euler-Heisenberg (E-H) effective Lagrangian \cite%
{Heisenberg} of QED truncated at any finite power of its Taylor expansion in
the field. This allows us to identify the self-coupling constant of the
electromagnetic field with a definite combination of the electron mass and
charge and to propose that such models may be used to extend QED to the
extreme distances smaller than those for which QED may be thought of as a
perfectly adequate theory.

More advanced approaches based on the Euler-Heisenberg Lagrangian that do
not depend upon any assumption of smallness of its field argument (the
background field) and do not hence appeal to expansion of the Lagrangian in
powers of the background fields received attention, as well, under the
restriction, however, that not-too-fast-varying in space and time fields are
studied as solutions of the nonlinear Maxwell equations. Among the nonlinear
effects studied, there are the linear and quadratic electric and magnetic
responses of the vacuum with a strong constant field in it to an applied
electric field \cite{AdoGitSha2016}, with the emphasis on the
magneto-electric effect \cite{GitSha2012, AdoGitSha2013} and
magnetic monopole formation \cite{AdoGitSha2015}. Also self-interaction of
electric and magnetic dipoles was considered with the indication that the
electric and magnetic moments of elementary particles are subjected to a
certain electromagnetic renormalization \cite{CosGitSha2013} after being
calculated following a strong-interaction theory, say, QCD or lattice
simulations. Interaction of two laser beams against the background of a slow
electromagnetic wave was studied along these lines, too \cite{King}. The
finiteness of the field energy allows one to develop a soliton view on a
moving point charge \cite{movingPreprint}, \cite{Shishmarev}.

In the present paper we are extending the consideration to cover the
electrostatic problem of a system of two point charges that interact
following nonlinear Maxwell equations stemming from the Lagrangian quadratic
in the the field invariant $\mathfrak{F}$. Their common field is not, of
course, just a linear combination of the individual fields of the two
charges. The nonlinear problem is outlined in the next Section \ref{NME},
where we present the nonlinear Maxwell equations and give them the form in
Subsection \ref{model} apt for finding the approximate solutions of Section %
\ref{Appr}. Once the field energy is finite it is possible in principle to
define the attraction or repulsion force between charges as the derivative
of the field energy with respect to the distance $\mathbf{R}$ between them.
Contrary to the standard linear electrodynamics, this is evidently not the
same as the product of one charge by the field strength produced by the
other! This rule holds true only if one of the charges is much smaller in
value than the other.

In Section \ref{Appr} we develop the procedure of finding the solution to
the static two-body problem in two opposite approximations determined by the
ratio of the distance $R,$ to the coordinate of the observation point $r$\footnote{
	Throughout the paper, Greek indices span Minkowski space-time, Roman indices
	span its three-dimensional subspace. Boldfaced letters are three-dimensional
	vectors, same letters without boldfacing and index designate their lengths,
	except the coordinate vector $\mathbf{x=r}$, whose length is denoted as $r.$
	The scalar product is ($\mathbf{r\cdot R)=}x_{i}R_{i},$ the vector product
	is $\mathbf{C=}\left[ \mathbf{r\times R}\right] ,$ $C_{i}=\epsilon
	_{ijk}x_{i}R_{k}$}.
Where this ratio is small, $R/r \ll 1$, we find in Subsection \ref{dipole} the leading expression for the common field, which makes the
nonlinear correction to electric dipole, and the corresponding potential. In
Subsection (\ref{largeSeparation})) the opposite case $r/R \ll 1$\ is
considered, first, also in the leading approximation (Subsubsection \ref{leading}). The simplifying circumstance that makes these
approximations easy to handle is that it so happens that one needs,
as a matter of fact, to solve only the second (following the
classification of Ref. \cite{Landau}) Maxwell equation$,$ the one
following from the least action principle, while the first one, the
Bianchi identity, $\left[ \mathbf{\nabla \times E}\right] =0,$ is
trivially satisfied. The situation becomes far more complicated in
the next-to-leading approximation $\left( {r}/{R}\right)^{2}$
developed in Subsubsection \ref{quadrupole}. In developing the above
approximations no assumption was made on whether the nonlinear scale
determined by the self-coupling of the electromagnetic field is
large or
small as compared to $r$ or $R.$ Their use in the expression for the energy of the two charges as a function of the separation $R$\ in the
limit $R\rightarrow 0,$ i.e.\ when the charges are so close to one another
that $R$ is much less than the nonlinearity scale, allows to make a
preliminary estimation confirmed by another approach to be reported in a
separate publication that for small separation, the energy of the system of
two point charges can be presented as $a+b R^{1/3},$ where $a$ and $b$
are finite constants depending only on the two charges (in QED they include
the electron mass and charge). Hence the force between two point-like
charges turns to infinity following the law $R^{-2/3}.$ This formula
replaces, in the given nonlinear model, the Coulomb law $R^{-2}$ for the
force between two point charges.

\section{Nonlinear Maxwell equations}
\label{NME}

\subsection{Nonlinear Maxwell equations as they originate from QED}
\label{QED}

It is known that QED is a nonlinear theory due to virtual electron-positron
pair creation by a photon. The nonlinear Maxwell equation of QED for the
electromagnetic field tensor $F_{\nu \mu }\left( x\right) =\partial ^{\mu
}A^{\nu }(x)-\partial ^{\nu }A^{\mu }(x)$ $(\tilde{F}_{\tau \mu }\left(
x\right) $ designates its dual tensor $\tilde{F}^{\mu \nu }=\left(
1/2\right) \varepsilon ^{\mu \nu \rho \sigma }F_{\rho \sigma }$ ) produced
by the classical source \ $J_{\mu }\left( x\right) $ may be written as, see
\textit{e.g.} \cite{AdoGitSha2016}.%
\begin{equation}
\partial ^{\nu }F_{\nu \mu }\left( x\right) -\partial ^{\tau }\left[ \frac{%
\delta \mathcal{L}\left( \mathfrak{F,G}\right) }{\delta \mathfrak{F}\left(
x\right) }F_{\tau \mu }\left( x\right) +\frac{\delta \mathcal{L}\left(
\mathfrak{F,G}\right) }{\delta \mathfrak{G}\left( x\right) }\tilde{F}_{\tau
\mu }\left( x\right) \right] =J_{\mu }\left( x\right) \,  \label{MaxEq}
\end{equation}%
Here $\mathcal{L}\left( \mathfrak{F,G}\right) $ is the effective Lagrangian
(a function of the two field invariants $\mathfrak{F}=(1/4)F^{\mu \nu
}F_{\mu \nu }$ and $\mathfrak{G}=\left( 1/4\right) \tilde{F}^{\mu \nu
}F_{\mu \nu }),$ of which the generating functional of
one-particle-irreducible vertex functions, called effective action \cite%
{weinberg}, is obtained by the space-time integration as $\textstyle \Gamma \left[ A%
\right] =\int \mathcal{L}\left( x\right) d^{4}x.$ Eq. (\ref{MaxEq}) is the
realization of the least action principle
\begin{equation}\nonumber
\frac{\delta S\left[ A\right] }{\delta A^{\mu }\left( x\right) }=\partial
^{\nu }F_{\nu \mu }\left( x\right) +\frac{\delta \Gamma \left[ A\right] }{%
\delta A^{\mu }\left( x\right) }=J_{\mu }(x),  %\label{nm3}
\end{equation}%
where the full action $S\left[ A\right] =S_{\mathrm{Maxw}}\left[ A\right]
+\Gamma \left[ A\right] $ includes the standard classical, Maxwellian,
electromagnetic action $S_{\mathrm{Maxw}}\left[ A\right] =-\int \mathfrak{F}%
\left( x\right) d^{4}x$ with its Lagrangian known as $L_{\text{Maxw}}=-%
\mathfrak{F}=(1/2)\left( E^{2}-B^{2}\right) $ in terms of the electric
and magnetic fields, $\mathbf{E}$ and $\mathbf{B}.$

Eq. (\ref{MaxEq}) is reliable only as long as its solutions vary but slowly
in the space-time variable $x_{\mu },$ because we do not include the space
and time derivatives of $\mathfrak{F}$ and$\mathfrak{\ G}$ as possible
arguments of the functional $\Gamma \left[ A\right] $ treated approximately
as local$.$ This infrared, or local approximation shows itself as a rather
productive tool \cite{AdoGitSha2016}-- \cite{King}. The calculation of one
electron-positron loop with\ the electron propagator taken as solution to
the Dirac equation in an arbitrary combination of constant electric and
magnetic fields of any magnitude supplies us with a useful example of $%
\Gamma \left[ A\right] $ known as the E-H effective action \cite{Heisenberg}. It is valid to the lowest order in the fine-structure constant $\alpha$, but with no restriction imposed on the the background field, except that it
has no nonzero space-time derivatives. Two-loop expression of this local functional is also available \cite{Ritus}.

The dynamical Eq. (\ref{MaxEq}), which makes the "second pair" of Maxwell
equations, may be completed by postulating also their "first pair"
\begin{equation}
\partial _{\nu }\widetilde{F}^{\nu \mu }\left( x\right) =0,
\label{bianchi}
\end{equation}%
whose fulfillment allows using the 4-vector potential $A^{\nu }(x)$ for
representation of the fields: $F_{\nu \mu }\left( x\right) =\partial ^{\mu
}A^{\nu }(x)-\partial ^{\nu }A^{\mu }(x).$ This representation is important
for formulating the least action principle and quantization of the
electromagnetic field. From it, Eq. (\ref{bianchi}) follows identically,
unless the potential has the angular singularity like the Dirac string
peculiar to magnetic monopole. In the present paper we keep to Eq. (\ref%
{bianchi}), although its local denial is not meaningless, as discussed in
Refs. \cite{AdoGitSha2015}, where a magnetic charge is produced in nonlinear
electrodynamics.

We are going now to separate the electrostatic case. This may be possible if
the reference frame exists where all the charges are at rest, $J_{0}(x) = J_{0}(\mathbf{r})$ (We denote $\mathbf{r=x}$). Then
in this "rest frame" the spacial component of the current disappears, $%
\mathbf{J}\left( x\right) =0,$ and the purely electric time-independent
configuration $F_{ij}\left( \mathbf{r}\right) =0$ would not contradict
equation (\ref{MaxEq}). With the magnetic field equal to zero, the invariant
$\mathfrak{G=}$ $\left( \mathbf{E}\cdot \mathbf{B}\right) $ disappears, too.
In a theory even under the space reflection, to which class QED belongs,
also we have $\textstyle \left. \frac{\partial \mathcal{L}\left( \mathfrak{F,G}\right)
}{\partial \mathfrak{G}\left( x\right) }\right\vert _{\mathfrak{G}=0}=0$, 
since the Lagrangian should be an even function of the pseudoscalar $\mathfrak{G}$. Then we are left with the equation for a static electric field $E_{i}=F_{i0}\left( \mathbf{x}\right)$%
\begin{equation}
\partial _{i}F_{i0}\left( \mathbf{r}\right) -\partial _{i}\frac{\delta
\mathcal{L}\left( \mathfrak{F,}0\right) }{\delta \mathfrak{F}\left( \mathbf{r%
}\right) }F_{i0}(\mathbf{r})=J_{0}\left( \mathbf{r}\right) .\,
\label{static}
\end{equation}

\subsection{Generalities of solutions to nonlinear Maxwell equations}
\label{model}
Equation (\ref{static}) is seen to be the equation of motion stemming
directly from the Lagrangian
\begin{equation}
L=-\mathfrak{F+}\mathcal{L}\left( \mathfrak{F,}0\right)  \label{L}
\end{equation}%
with the constant external charge $J_{0}\left( \mathbf{r}\right) $ and the
zero argument set for the second field invariant $\mathfrak{G}.$ In the rest
of the paper we shall be basing on this Lagrangian in understanding that it
may originate from QED as described above or, alternatively, be given
\textit{ad hoc} to define a certain model\textit{. }In the latter case, if
treated seriously as applied to short distances near a point charge where
the field cannot be considered as slowly varying, in other words, beyond the
applicability of the infrared approximation of QED outlined above, the
Lagrangian (\ref{L}) may be referred to as defining an extension of QED to
short distances once $\mathcal{L}\left( \mathfrak{F,}0\right) $ is the E-H
Lagrangian (or else its multi-loop specification) restricted to $\mathfrak{G}%
=0$.

It was shown in \cite{CosGitSha2013a} that the important property of
finiteness of the field energy of the point charge is guarantied if $%
\mathcal{L}\left( \mathfrak{F,}0\right) $ in (\ref{L}) is a polynomial of
any power, obtained, for instance, by truncating the Taylor expansion of the
H-E Lagrangian at any integer power of $\mathfrak{F}$. On the other hand, it
was indicated in \cite{movingPreprint} that a weaker condition is
sufficient: if $\mathcal{L}\left( \mathfrak{F,}0\right) \mathfrak{\ }$grows
with $-\mathfrak{F}$ as $\left( -\mathfrak{F}\right) ^{w}$, the field energy
is finite provided that $w>3/2$. The derivation of this condition is
given in \cite{AdoGitShaShi2016} and in \cite{Ependiev}. As a matter of fact
a more subtle condition suffices: $\mathcal{L}\left( \mathfrak{F}\right)
\sim \left( -\mathfrak{F}\right) ^{\frac{3}{2}}\ln ^{u}\left( -\mathfrak{F}%
\right) ,$ \ $u>2.$

In the present paper we confine ourselves to the simplest example of the
nonlinearity generated by keeping only quadratic terms in the Taylor
expansion of the E-H Lagrangian in powers of the field invariant $\mathfrak{F%
}$
\begin{equation*}
\mathcal{L}\left( \mathfrak{F((}x)\mathfrak{,}0\right) =\frac{1}{2}\left.
\frac{d^{2}\mathcal{L}\left( \mathfrak{F,}0\right) }{d^{2}\mathfrak{F}}%
\right\vert _{\mathfrak{F}=0}\mathfrak{F}^{2}\mathfrak{(}x),
\end{equation*}%
where the constant and linear terms are not kept, because their inclusion
would contradict the correspondance principle that does not admit changing
the Maxwell Lagrangian $L_{\text{Max}}=-\mathfrak{F}$ for small fields. The
correspondence principle is laid into the calculation of the E-H Lagrangian
via the renormalization procedure.

Finally, we shall be dealing with the model Lagrangian quartic in the field
strength%
\begin{equation}
L=-\mathfrak{F(}x)\mathfrak{+}\frac{1}{2}\gamma \mathfrak{F}^{2}\mathfrak{(}%
x)  \label{quartic}
\end{equation}%
with $\gamma $ being a certain self-coupling coefficient with the
dimensionality of the fourth power of the length, which may be taken as
\begin{equation}
\gamma =\left. \frac{d^{2}\mathcal{L}\left( \mathfrak{F,}0\right) }{d^{2}%
\mathfrak{F}}\right\vert _{\mathfrak{F}=0}=\frac{e^{4}}{45\pi ^{2}m^{4}},
\label{gamma}
\end{equation}%
where $e$ and $m$ are the charge and mass of the electron, if $\mathcal{L}$
is chosen to be the E-H one-loop Lagrangian. We do not refer to this choice
henceforward. Generalization to general Lagrangians can be also done in a
straightforward way.

The second (\ref{static}) and the first (\ref{bianchi}) Maxwell equations
for the electric field $\mathbf{E}$ with Lagrangian (\ref{quartic}) are
\begin{gather}
\mathbf{\nabla }\cdot \left[ \left( 1+\frac{\gamma }{2}E^{2}(\mathbf{r}%
)\right) \mathbf{E}(\mathbf{r})\right] =j_{0}(\mathbf{r}),  \label{E_div} \\
\mathbf{\nabla }\times \mathbf{E}(\mathbf{r})=0.  \label{rot}
\end{gather}%
Denoting the solution of the linear Maxwell equations as $\mathbf{E}^{lin}(%
\mathbf{r})$
\begin{equation}\nonumber
\nabla \cdot \mathbf{E}^{lin}(\mathbf{r})=j_{0}(\mathbf{r}),\quad \mathbf{%
\nabla }\times \mathbf{E}^{lin}(\mathbf{r})=0,  %\label{eqM_lin}
\end{equation}%
we write the solution of (\ref{E_div}), in the following way \cite%
{AdoGitSha2016} -- \cite{CosGitSha2013}%
\begin{equation}
\left( 1+\frac{\gamma }{2}E^{2}(\mathbf{r})\right) \mathbf{E}(\mathbf{r})=%
\mathbf{E}^{lin}(\mathbf{r})+[\mathbf{\nabla }\times \mathbf{\Omega }(%
\mathbf{r})],  \label{E_alg}
\end{equation}%
because $\nabla \cdot $\ $[\mathbf{\nabla }\times \mathbf{\Omega }(\mathbf{r}%
)]=0.$

The second Maxwell equation (\ref{E_alg}) may be conveniently written in the
form to be exploited later
\begin{equation}
\mathbf{E}(\mathbf{r})=\mathbf{N}(\mathbf{r})\xi \left( \gamma N^{2}(\mathbf{%
r})\right) ,\quad \mathbf{N}(\mathbf{r})=\mathbf{E}^{lin}(\mathbf{r})+\nabla
\times \mathbf{\Omega }(\mathbf{r}),  \label{vExi}
\end{equation}%
where the function $\xi (x)$ is defined as the real solution to the cubic
equation
\begin{equation}
\left( 1+\frac{x}{2}\xi ^{2}(x)\right) \xi (x)=1,\quad x\geq 0.  \label{ksi}
\end{equation}%
Its explicit form is given by the Cardano formula:
\begin{equation}
\xi (x)=x^{-1/3}\left( \left[ \sqrt{1+\frac{8}{27x}}+1\right] ^{1/3}-\left[
\sqrt{1+\frac{8}{27x}}-1\right] ^{1/3}\right) .  \label{ksi1}
\end{equation}%
We substitute (\ref{vExi}) in the first Maxwell equation (\ref{rot}) to get:
\begin{equation}%\nonumber
\nabla \times (\mathbf{N}\xi (\gamma N^{2}))=\xi (\gamma N^{2})[\nabla
\times \mathbf{N}]-\gamma \xi ^{\prime }(\gamma N^{2})[\mathbf{N}\times
\nabla ]N^{2})=0,  \label{eqNNN2}
\end{equation}%
where the prime designates the derivative with respect to the argument.
Taking into account the relations
\begin{gather}
\nabla \times \mathbf{N}=\nabla \times \mathbf{E}^{lin}+\nabla \times
\lbrack \nabla \times \mathbf{\Omega }]=\nabla (\nabla \cdot \mathbf{\Omega }%
)-\Delta \mathbf{\Omega }=-\Delta \mathbf{\Omega },  \notag \\
\nabla N^{2}=2\left( \mathbf{N}\times \lbrack \nabla \times \mathbf{N}]+(%
\mathbf{N}\cdot \vec{\nabla})\mathbf{N}\right) =2\left( -\mathbf{N}\times
\lbrack \Delta \mathbf{\Omega }]+(\mathbf{N}\cdot \vec{\nabla})\mathbf{N}%
\right) ,  \notag \\
\mathbf{N}\times (\mathbf{N}\times \lbrack \Delta \mathbf{\Omega }%
])=-N^{2}[\Delta \mathbf{\Omega }],  \notag \\
\xi ^{\prime }(x)=-\frac{\xi ^{3}(x)}{2+3x\xi ^{2}(x)}=-\frac{\xi (x)(\xi
(x)-1)}{x(2\xi (x)-3)},  \notag
\end{gather}%
for (\ref{eqNNN2}) we obtain
\begin{equation}
-\frac{\xi (\gamma N^{2})}{3-2\xi (\gamma N^{2})}\left( \Delta \mathbf{%
\Omega }-\frac{2}{N^{2}}(1-\xi (\gamma N^{2}))\mathbf{N}\times \lbrack (%
\mathbf{N}\cdot \vec{\nabla})\mathbf{N}]\right) =0.
\end{equation}%
Since function $\xi(x)/(3-2\xi (x))$ does not have zeros on $%
[0;\infty )$, equation (\ref{eqNNN2}) is equivalent to the equation:
\begin{equation}
\Delta \mathbf{\Omega }=f(\mathbf{\Omega }),\quad f(\mathbf{\Omega })=\frac{2%
}{N^{2}}(1-\xi (\gamma N^{2}))\mathbf{N}\times \lbrack (\mathbf{N}\cdot \vec{%
\nabla})\mathbf{N}].  \label{eqOm}
\end{equation}

In the center-symmetric case of a single point charge considered in \cite%
{CosGitSha2013a}, \cite{movingPreprint}, \cite{AdoGitShaShi2016}, \cite%
{Shishmarev}, one has $\mathbf{\Omega }(\mathbf{r})=0,$ as a solution to the
equation (\ref{eqOm}). This simplification makes the exact solution
possible. The equality $\mathbf{\Omega }(\mathbf{r})=0$ holds as well in the
axial-symmetric problem of two point charges within the approximations
linear with respect to the ratios $R/r$ or $r/R$ to be
considered in Subsections \ref{dipole} and \ref{largeSeparation} of the
next Section. In these cases it will be sufficient to present the solution
of the differential part of Eq. (\ref{E_div}) in the form (\ref{E_alg})
setting $\mathbf{\Omega }(\mathbf{r})=0$ in it, then the first Maxwell
equation (\ref{rot}) is fulfilled automatically. On the contrary, within the
next order of $\textstyle \left( r/R\right)^{2}$ the pseudovector function $%
\mathbf{\Omega }(\mathbf{r})$ is nontrivial, which makes the
axial-symmetric "quadrupole-like" solution found in Subsection
\ref{quadrupole} for the field of two point-like charges more
sophisticated.

\section{Two-body problem}
\label{Appr}
By the two point-charge problem we mean the one, where the current $j_{0}(%
\mathbf{r})$ in (\ref{E_div}) is the sum of delta-functions centered in the
positions $\mathbf{r=\pm R}$ of two charges $q_{1}$ and $q_{2}$ separated by
the distance $2R$ (with the origin of coordinates $x_{i}$ placed in the
middle between the charges)%
\begin{equation}
\mathbf{\nabla }\cdot \left[ \left( 1+\frac{\gamma }{2}E^{2}(\mathbf{r}%
)\right) \mathbf{E}(\mathbf{r})\right] =q_{1}\delta ^{3}\left( \mathbf{r-R}%
\right) +q_{2}\delta ^{3}\left( \mathbf{r}+\mathbf{R}\right) .
\label{two-body}
\end{equation}%
In what follows we shall be addressing this equation as accompanied by (\ref%
{rot}) for the combined field of two charges.

In what follows we shall refer to the field energy density that in the
present model (\ref{quartic}), when there is electric field alone, reads%
\begin{equation}
\Theta ^{00}=(1+\frac{\gamma E^{2}}{2})E^{2}-\frac{E^{2}}{2}\left( 1+\frac{%
\gamma E^{2}}{4}\right) =\frac{E^{2}}{2}+\frac{3\gamma E^{4}}{8}.
\label{theta}
\end{equation}

The integral for the full energy of two charges
\begin{equation}
P^{0}=\int \Theta ^{00}d^{3}x  \label{P0}
\end{equation}
converges since it might diverge only when integrating over close vicinities
of the charges. But in each vicinity the field of the nearest charge
dominates, and we know from the previous publication \cite{CosGitSha2013a}
(also to be explained below) that the energy of a separate charge converges
in the present model. When the charges are in the same point, $R=0,$ they
make one charge $q_{1}+q_{2},$ whose energy converges, too.

\subsection{Small separation $r\gg R$ between charges (dipole approximation)}
\label{dipole}
We shall be looking for the solution of (\ref{two-body}) in the
form
\begin{equation*}
\mathbf{E}=\mathbf{E}^{(0)}+\mathbf{E}^{(1)}+...
\end{equation*}%
where $\mathbf{E}^{(0)}$ and $\mathbf{E}^{(1)}$ are contributions of the
zeroth and first order with respect to the ratio $R/r\ll 1,$
respectively. This strong inequality means that the observation point is
far from the location of the both charges. So the result of consideration in
the present subsection will be the extension of the dipole field to the
case, where the point charges self-interact and interact nonlinearly with
each other.

The zero-order term is spherical-symmetric, because it corresponds to two
charges $q_{1}$, $q_{2}$ in the same point that make one charge $q_{1}+q_{2}$,
\begin{equation}
\mathbf{E}^{(0)}(\mathbf{r})=\frac{\mathbf{r}}{r}E^{(0)}(r)=\frac{\mathbf{r}%
}{r}\frac{q_{1}+q_{2}}{4\pi r^{2}}\xi \left( \gamma \left( \frac{q_{1}+q_{2}%
}{4\pi r^{2}}\right) ^{2}\right) ,  \label{anz1}
\end{equation}%
where $\xi(x)$ is the solution (\ref{ksi1}) of equation (\ref{ksi1}). Eq.
(\ref{rot}) is automatically fulfilled by the centre-symmetric form (\ref%
{anz1}). The field $\mathbf{E}^{(0)}$ is the nonlinear extension \cite%
{CosGitSha2013a}\ of the standard Coulomb field 
\begin{equation}\nonumber
	\frac{q_{1}+q_{2}}{4\pi r^{2}}\frac{\mathbf{r}}{r}
\end{equation}
of the sum charge.

Let us write the first-order term $E_{i}^{(1)}$ in the following general
axial-symmetric form, linear in the ratio $\mathbf{R}/r$:
\begin{equation}
\mathbf{E}^{(1)}=\mathbf{r}\left( \mathbf{R\cdot r}\right) a(r)+\mathbf{R}%
g(r),  \label{cylform}
\end{equation}%
where $a$ and $g$ \ are functions of the only scalar $r,$ and the symmetry
axis is fixed as the line passing through the two charges. Let us subject (%
\ref{cylform}) to the equation (\ref{rot}) $\nabla \times \mathbf{E}%
^{(1)}=0. $ This results in the relation%
\begin{equation}
a(r)=\frac{1}{r}\frac{\text{d}}{\text{d}r}g(r),  \label{connection}
\end{equation}%
provided that the vectors $\mathbf{r,R}$ are not parallel. We shall see that
with the ansatzes (\ref{cylform}) and (\ref{anz1}) equation (\ref{E_alg})
can be satisfied with the choice $\mathbf{\Omega }(\mathbf{r})=0$:
\begin{equation}
\left( 1+\frac{\gamma }{2}E^{2}(\mathbf{r})\right) \mathbf{E}(\mathbf{r})=%
\mathbf{E}^{lin}(\mathbf{r}),  \label{without}
\end{equation}%
namely, we shall find the coefficient functions $a,$ $g$ from Eq. (\ref%
{without}) and then ascertain that the relation ( \ref{connection}) is
obeyed by the solution.

The inhomogeneity in (\ref{without})%
\begin{equation}
\mathbf{E}^{lin}\mathbf{(r})=\frac{q_{1}}{4\pi }\frac{\mathbf{r-R}}{|\mathbf{%
r-R}|^{3}}+\frac{q_{2}}{4\pi }\frac{\mathbf{r+R}}{|\mathbf{r+R}|^{3}}
\label{inhomogeneity}
\end{equation}%
satisfies the linear ($\gamma =0)$ limit of equation (\ref{two-body})
\begin{equation}\nonumber
\mathbf{\nabla }\cdot \mathbf{E}^{lin}(\mathbf{r})=q_{1}\delta ^{3}\left(
\mathbf{r-R}\right) +q_{2}\delta ^{3}\left( \mathbf{r}+\mathbf{R}\right)
\end{equation}%
and also (\ref{rot}). The inhomogeneity (\ref{inhomogeneity}) is expanded in
$\mathbf{R}/r$ as
\begin{equation}
\mathbf{E}^{lin}\mathbf{(r})=\frac{(q_{1}+q_{2})}{4\pi r^{2}}\frac{\mathbf{r}%
}{r}+\frac{\left( q_{2}-q_{1}\right) }{4\pi r^{2}}\left( \frac{\mathbf{R}}{r}%
-3\frac{\mathbf{r}}{r}\frac{\left( \mathbf{R\cdot r}\right) }{r^{2}}\right)
+...  \label{lindip}
\end{equation}%
This is the standard monopole+dipole approximation in understanding that $%
\mathbf{d=}$ $\left( q_{2}-q_{1}\right) \mathbf{R}$ is the dipole moment of
the two charges, while the dots stand for the disregarded quadrupole and
higher multipole contributions.

The zero-order term satisfies the equation%
\begin{equation}
\left( 1+\frac{\gamma }{2}E^{(0)2}(r)\right) E^{(0)}(r)=\frac{q_{1}+q_{2}}{
4\pi r^{2}},  \label{zeroord}
\end{equation}%
with the first term of expansion (\ref{inhom}) taken for inhomogeneity. This
is an algebraic (not differential) equation, cubic in the present model (\ref%
{quartic}), solved explicitly for the field $E^{(0)}$ as a function of $r$
in this case, but readily solved for the inverse function $r(E^{(0)})$ in
any model$,$ this solution being sufficient for many purposes. Even without
solving it we see that for small $r\ll \gamma ^{1/4}$ the second
term in the bracket dominates over the unity, therefore the asymptotic
behavior in this region follows from (\ref{zeroord}) to be%
\begin{equation}\nonumber
E^{(0)}(r)\sim \left( \frac{q_{1}+q_{2}}{2\pi \gamma }\right) ^{\frac{1}{3}%
}r^{-\frac{2}{3}}.  %\label{E(0)}
\end{equation}%
This weakened -- as compared to the Coulomb field $(q_1+q_2)/(4\pi)r^{-2}$ -- singularity is not an obstacle to convergence of the both
integrals in (\ref{P0}) for the proper field energy of the equivalent point
charge $q_{1}+q_{2}.$

With the zero-order equation (\ref{zeroord}) fulfilled, we write a linear
algebraic equation for the first-order correction $\mathbf{E}^{(1)}$ from (%
\ref{without}), to which the second, dipole part in (\ref{inhom}) serves as
an inhomogeneity
\begin{equation}
\mathbf{E}^{(1)}=\frac{q_{2}-q_{1}}{4\pi r^{2}}\left( \frac{%
\mathbf{R}}{r}-3\frac{\mathbf{r}}{r}\frac{\left( \mathbf{R\cdot r}\right) }{%
r^{2}}\right) -\frac{\gamma }{2}\left[ 2\left( \mathbf{E}^{(1)}\cdot \mathbf{%
E}^{(0)}\right) \mathbf{E}^{(0)}+E^{(0)2}\mathbf{E}^{(1)}\right] .
\label{E1}
\end{equation}%
This equation is linear and it does not contain derivatives. We use (\ref%
{cylform}) as the ansatz. After calculating%
\begin{equation*}
2\left( \mathbf{E}^{(1)}\cdot \mathbf{E}^{(0)}\right) \mathbf{E}%
^{(0)}+E^{(0)2}\mathbf{E}^{(1)}=\mathbf{r}E^{(0)2}\frac{\left( \mathbf{%
R\cdot r}\right) }{r^{2}}\left( 2g+3r^{2}a\right) +\mathbf{R}gE^{(0)2},
\end{equation*}%
we obtain two equations, along $\mathbf{R}$ and $\mathbf{r,}$ with the
solutions:
\begin{equation}
g=\frac{\delta q}{r^{3}}\frac{1}{1+\frac{\gamma }{2}E^{(0)2}}=\frac{\delta q%
}{Qr}E^{(0)},  \label{gg}
\end{equation}

\begin{equation}
a=-\frac{\delta q}{r^{5}}\frac{3+\frac{5\gamma }{2}E^{(0)2}}{\left( 1+\frac{%
\gamma }{2}E^{(0)2}\right) \left( 1+\frac{3\gamma }{2}E^{(0)2}\right) },
\label{a}
\end{equation}
where $\delta q=(q_{2}-q_{1})/(4\pi)$, $Q=(q_{2}+q_{1})/(4\pi)$. From (\ref{zeroord}) we obtain
\begin{equation*}
\frac{\text{d}}{\text{d}r}E^{(0)}=-\frac{2Q}{r^{3}\left( 1+\frac{\gamma }{2}%
E^{(0)2}\right) }-\frac{\gamma E^{(0)2}}{1+\frac{\gamma }{2}E^{(0)2}}\frac{%
\text{d}}{\text{d}r}E^{(0)}.
\end{equation*}%
Hence
\begin{equation}\nonumber
\frac{\text{d}}{\text{d}r}E^{(0)}=-\frac{2Q}{r^{3}\left( 1+\frac{3\gamma }{2}%
E^{(0)2}\right) }.  %\label{dE/dr}
\end{equation}%
With the help of this relation the derivative of (\ref{gg}) can be
calculated to coincide with $\left( \ref{a}\right) $ times $r.$ This proves
Eq. (\ref{connection}) necessary to satisfy the first Maxwell equation (\ref%
{rot}).

By substituting relations (\ref{gg}) and (\ref{a}) in the decomposition (\ref%
{cylform}) we finally have%
\begin{equation}
\mathbf{E}=\frac{\mathbf{r}}{r}E^{(0)}(r)+\frac{q_{2}-q_{1}}{4\pi \text{ }%
\left( 1+\frac{\gamma }{2}E^{(0)2}\right) }\left[ \frac{\mathbf{R}}{r^{3}}-%
\frac{\mathbf{r}\left( \mathbf{R\cdot r}\right) }{r^{5}}\frac{3+\frac{%
5\gamma }{2}E^{(0)2}}{1+\frac{3\gamma }{2}E^{(0)2}}\right]  \label{result1}
\end{equation}%
for the solution of the both Maxwell equations up to $O(R^2/r^2)$.

The potential corresponding to the electric field (\ref{result1}) has the
form
\begin{equation}
\varphi =V_{0}(r)+\frac{q_{1}-q_{2}}{q_{1}+q_{2}}\frac{E^{(0)}(r)}{r}(%
\mathbf{r}\cdot \mathbf{R}),  \label{pot1}
\end{equation}%
where $V_{0}(r)$ is the potential of the field of one charge \cite%
{AdoGitShaShi2016}:
\begin{gather}\nonumber
V_{0}(r)=\int_{r}^{\infty }E^{(0)}(r)dr= \\ \nonumber
-rE^{(0)}(r)+\operatorname{sign}(Q)\sqrt{|Q|}\left( \frac{2}{\gamma }\right) ^{1/4}%
\mathcal{F}\left[ 2\arctan \left( \sqrt{\frac{\gamma }{2}}%
|E^{(0)}(r)|\right) ^{1/2},\frac{1}{2}\right] ,  \notag
\end{gather}%
where $\mathcal{F}(\phi ,m)=\int_{0}^{\phi }(1-m\sin ^{2}\theta
)^{-1/2}d\theta $ is elliptic integral of the first kind.

The first term (\ref{inhom}) is the nonlinear electric monopole field (\ref%
{anz1}) substituting for the Coulomb field in the nonlinear problem
under study, while the second term in (\ref{inhom}) may be
considered as giving nonlinear correction to the electric dipole
field . The lines of force and the equipotential curves of the
latter field drawn under the choice of parameters corresponding to a
strong nonlinearity are shown in Fig. 1

\begin{figure}[h]
\center{\includegraphics[scale=1.4]{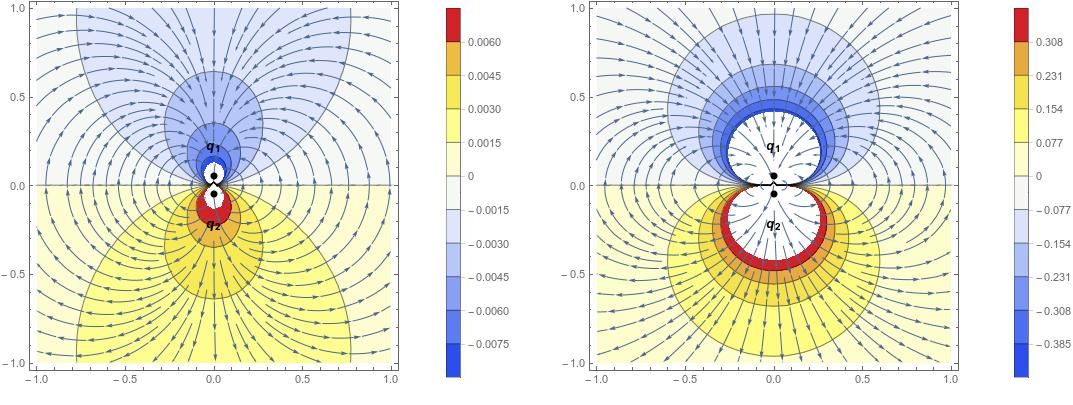}} \label{dipoleField}
\caption{The left graph corresponds to the nonlinear dipole field, second
term in Eq. (\protect\ref{result1}); the right graph shows the standard
dipole field. We use $R=0.05$, $q_{1}=-10$, $q_{2}=8$, $\protect\gamma %
^{1/4}=40$. Besides the vast difference in the magnitude scale, the left
pattern shows flattening towards the vertical axis, on which two bold dots
indicating two charges are placed.}
\end{figure}

\subsection{Large separation $r\ll R$ between charges}
\label{largeSeparation}

The quantities that relate to the approximation valid at $r\ll R,$ dealt
with in this Subsection, will be supplied with tilde to be distinguished
from the corresponding quantities in the previous Subsection \ref{dipole}
relating to the opposite approximation.

Let us expand the inhomogeneity (\ref{inhomogeneity}) to the first order in
the ratio $\mathbf{r}/R$ (without assuming the smallness of $R$ and $r$ as compared to $\gamma^{1/4})$:
\begin{equation}
\mathbf{E}^{lin}\mathbf{(r})=\frac{q_{2}-q_{1}}{4\pi R^{2}}\frac{\mathbf{R}}{%
R}+\frac{\left( q_{2}+q_{1}\right) }{4\pi R^{2}}\left( \frac{\mathbf{r}}{R}-%
\frac{3\mathbf{R}}{R}\frac{\left( \mathbf{R\cdot r}\right) }{R^{2}}\right)
+\dots  \label{inhom}
\end{equation}%
The first term here has a clear meaning of the sum of two oppositely
directed Coulomb fields produced in the point $r=0$\ by the two charges
placed far from one another$.$ The second one looks like a dipole field in
the variable $R$\ with the equivalent "dipole moment" $\left(
q_{2}+q_{1}\right) \mathbf{r.}$

We are looking for solution to equation (\ref{without}) in the form of the
expansion in powers of $\mathbf{r}/R$
\begin{equation}
\mathbf{E}=\widetilde{\mathbf{E}}^{(0)}+\widetilde{\mathbf{E}}^{(1)}+\dots=%
\frac{f}{4\pi R^{2}}\frac{\mathbf{R}}{R}+\frac{1}{4\pi R^{2}}\left( \frac{%
\mathbf{r}}{R}c+\frac{\mathbf{R}}{R}\frac{\left( \mathbf{R\cdot r}\right) }{%
R^{2}}b\right) +\dots  \label{E+E}
\end{equation}%
with the the yet unknown dimensionless coefficients $f,$ $c$ and $b$ being
functions of $R.$ The first Maxwell equation $\left[ \mathbf{\nabla \times E}%
\right] =0$ is satisfied by (\ref{E+E}).

In the zeroth order we have the equation for 
\begin{equation}\nonumber
\widetilde{\mathbf{E}}%
^{(0)}(R)=\frac{f}{4\pi R^{2}}\frac{\mathbf{R}}{R}=\widetilde{E}^{(0)}(R)%
\frac{\mathbf{R}}{R}
\end{equation}
in form
\begin{equation}
\widetilde{E}^{(0)}\left( 1+\frac{\gamma }{2}\widetilde{E}^{(0)2}\right) =%
\frac{q_{2}-q_{1}}{4\pi R^{2}},  \label{tldeE0}
\end{equation}%
which implies that $f/(4\pi R^{2})=\widetilde{E}^{(0)}(R)$ is the
function obtained from $E^{(0)}(r)$ of the previous Subsection \ref{dipole} by the substitution $r\rightarrow R$ and $q_{2}+q_{1}\rightarrow q_{2}-q_{1}$.

\subsubsection{Leading (dipole-like) approximation\label{leading}}

In the first order, the use of (\ref{tldeE0}) turns equation (\ref{without})
to the linear algebraic equation for $\widetilde{\mathbf{E}}^{(1)}(R)$
\begin{equation}
\widetilde{\mathbf{E}}^{(1)}=\frac{\left( q_{2}+q_{1}\right) }{4\pi R^{2}}%
\left( \frac{\mathbf{r}}{R}-\frac{3\mathbf{R}}{R}\frac{\left( \mathbf{R\cdot
r}\right) }{R^{2}}\right) -\frac{\gamma }{2}\left[ 2\left( \widetilde{%
\mathbf{E}}^{(1)}\cdot \widetilde{\mathbf{E}}^{(0)}\right) \widetilde{%
\mathbf{E}}^{(0)}+\widetilde{E}^{(0)2}\widetilde{\mathbf{E}}^{(1)}\right],
\label{E1eq}
\end{equation}%
Calculating the second term in the right-hand side (the auxiliary electric
field $\boldsymbol{\mathcal{E}}(\mathbf{r}) = (\gamma/2) E^2(\mathbf r)\mathbf{E}(\mathbf{r})$) with the ansatz (\ref{E+E})%
\begin{equation*}
2\left( \widetilde{\mathbf{E}}^{(1)}\cdot \widetilde{\mathbf{E}}%
^{(0)}\right) \widetilde{\mathbf{E}}^{(0)}+\widetilde{\mathbf{E}}^{(0)2}%
\widetilde{\mathbf{E}}^{(1)}=\frac{\mathbf{R}}{R}\widetilde{E}^{(0)2}\frac{%
\left( \mathbf{R\cdot r}\right) }{4\pi R^{4}}\left[ 2c+3b\right] +\mathbf{r}%
\widetilde{E}^{(0)2}\frac{c}{4\pi R^{3}},
\end{equation*}%
we obtain from (\ref{E1eq}) two equations for the components of $\mathbf{E}%
^{(1)}$ along $\mathbf{R}$ and along $\mathbf{r}$ that determine the values
\begin{equation}\nonumber
	c = \frac{\left( q_{2}+q_{1}\right) }{1+\frac{\gamma }{2}\widetilde{E}^{(0)2}}=
	\frac{q_{2}+q_{1}}{q_{2}-q_{1}}\widetilde{E}^{(0)}R^{2}, \quad
	b =-c\frac{3+\frac{5\gamma }{2}\widetilde{E}^{(0)2}}{1+\frac{3\gamma }{2}\widetilde{E}^{(0)2}}.
\end{equation}
Finally
\begin{eqnarray}
\mathbf{E} &=&\widetilde{\mathbf{E}}^{(0)}+\widetilde{\mathbf{E}}^{(1)}=
\notag \\
&=&\widetilde{E}^{(0)}(R)\frac{\mathbf{R}}{R}+\frac{q_{2}+q_{1}}{4\pi
R^{2}\left( 1+\frac{\gamma }{2}\widetilde{E}^{(0)2}(R)\right) }\left( \frac{%
\mathbf{r}}{R}-\frac{\mathbf{R}}{R}\frac{\left( \mathbf{R\cdot r}\right) }{%
R^{2}}\frac{3+\frac{5\gamma }{2}\widetilde{E}^{(0)2}(R)}{1+\frac{3\gamma }{2}%
\widetilde{E}^{(0)2}(R)}\right) ,  \label{result}
\end{eqnarray}%
where $\widetilde{E}^{(0)}$ is the solution of equation (\ref{tldeE0}) as a function of $R.$ The field (\ref{result}) obviously satisfies
the first Maxwell equation $\left[ \mathbf{\nabla \times E}\right] =0.$ By
comparing (\ref{result}) with the linear field of two charges in the similar
approximation (\ref{inhom}) we observe that in the zero-order term, the
difference $(q_{2}-q_{1})R^{-2}$ of the two Coulomb fields in the
point $r=0$ has been replaced by the nonlinear field $\widetilde{E}^{(0)}$
of the equivalent charge $q_{2}-q_{1}$, while in the first-order term the
"dipole field" $\left( q_{2}+q_{1}\right) \mathbf{r}$\ has been modified by
two different factors in the terms parallel to $\mathbf{r}$\ \ and $\mathbf{R%
}.$

To be more general, note that the fields (\ref{result}) and (\ref{result1})
turn into one another under the simultaneous replacement of the observation
coordinate $r$ by the separation $R$ between charges, and of the sum $%
q_{2}+q_{1}$ of the charges by their difference $q_{2}-q_{1}.$ The same
symmetry under the interchange $r\leftrightarrow R,$ $q_{2}+q_{1}%
\leftrightarrow q_{2}-q_{1}$ certainly holds for the linear $\gamma =0$
limits (\ref{inhom})$,$(\ref{lindip}) of Eqs. (\ref{result}) and (\ref%
{result1}). This symmetry occurs, because the second Maxwell equations
within the approximations adopted in this Subsubsection, $r\ll R,$ (\ref%
{E1eq}), and in the previous Subsection, $r\gg R,$ Eq. (\ref{E1}), turn into
each other under the transformation under consideration, while the first
Maxwell equation is satisfied for the both. As for the exact equation (\ref%
{without}), this transformation maps it into a strange differential equation
of a nonexisting theory.

\subsubsection{Next-to-leading (quadrupole-like) approximation}
\label{quadrupole}

In this Subsubsection we are studying the next, quadratic in the ratio $r/R$, term $\widetilde{\mathbf{E}}^{(2)}$\ extending the expansion (%
\ref{E+E}). To this end we first extend the expansion (\ref{inhom}) of the
linear field (\ref{inhomogeneity}) $\mathbf{E}^{lin}(\mathbf{r})$ to include
the corresponding term:
\begin{gather}
\mathbf{E}^{lin}(\mathbf{r})=\frac{q_{2}-q_{1}}{4\pi R^{2}}\frac{\mathbf{R}}{%
R}+\frac{q_{1}+q_{2}}{4\pi R^{2}}\left( \frac{\mathbf{r}}{R}-3\frac{\mathbf{R%
}}{R}\frac{(\mathbf{r}\cdot \mathbf{R})}{R^{2}}\right) +  \notag \\
+\frac{3(q_{1}-q_{2})}{8\pi R^{2}}\left( 2\frac{(\mathbf{r}\cdot \mathbf{R})%
}{R^{2}}\frac{\mathbf{r}}{R}+\left( \frac{r}{R}\right) ^{2}\frac{\mathbf{R}}{%
R}-5\frac{(\mathbf{r}\cdot \mathbf{R})^{2}}{R^{4}}\frac{\mathbf{R}}{R}%
\right) +O\left( \frac{r^{3}}{R^{3}}\right).  \label{lin-quadrupole}
\end{gather}
Once, up to the first order in $r/R$, Eq. (\ref{E+E}) satisfies the
first Maxwell equation $\left[ \mathbf{\nabla \times E}\right] =0$
automatically with any coefficients $f,b,c$ we conclude, as we did in the
previous Subsection, that the curl $[\mathbf{\nabla }\times \mathbf{\Omega }(%
\mathbf{r})]$ involved in (\ref{E_alg}) is zero to this order, Eq. (\ref%
{result}) being the solution to equation (\ref{E_alg}) without this curl.
This implies that the expansion of $[\mathbf{\nabla }\times \mathbf{\Omega }(%
\mathbf{r})]$ starts with the quadratic term $(r/R)^{2}$. Bearing in mind that the vector product $\left[ \mathbf{r}\times\mathbf{R}\right] $ is the only pseudovector in our problem\ and that the
action of $\mathbf{\nabla }$ lowers the power of $r$ by one we look for the
pseudovector $\mathbf{\Omega }$ in the form
\begin{equation}
\mathbf{\Omega =}\frac{\left[ \mathbf{r}\times \mathbf{R}\right] }{|\left[
\mathbf{r}\times \mathbf{R}\right] |}\left[ \Omega _{\phi }\left( \frac{%
\mathbf{r}\cdot \mathbf{R}}{rR}\right) \frac{r^{3}}{R^{3}}+O\left( \frac{%
r^{4}}{R^{4}}\right) \right] ,  \label{pseudo}
\end{equation}%
where $\Omega _{\phi }\left( (\mathbf{r}\cdot \mathbf{R})/(r R)\right)$
is a scalar function of the angle $\theta $ between the observation
direction and the axis, on which the charges lie, $\cos \theta =(\mathbf{r}\cdot \mathbf{R})/(r R)$. The straightforward calculation yields (we
refer to the orts $\mathbf{e}_{r}=\frac{\mathbf{r}}{r},\quad \mathbf{e}%
_{\phi }=-\frac{\mathbf{r}\times \mathbf{R}}{|\mathbf{r}\times \mathbf{R}|}%
,\quad \mathbf{e}_{\theta }=\mathbf{e}_{\phi }\times \mathbf{e}_{r}$ \ and
to the relation $\left( \nabla \cdot \mathbf{\Omega }\right) =0)$ obeyed by (%
\ref{pseudo}))
\begin{gather}
\left[ \nabla \times \mathbf{\Omega }\right] =\frac{1}{R}\left[ (\cot \theta
\Omega _{\phi }(\theta )+\frac{d\Omega _{\phi }}{d\theta })\mathbf{e}%
_{r}-4\Omega _{\phi }(\theta )\mathbf{e}_{\theta }\right] \left( \frac{r}{R}%
\right) ^{2},  \label{nablacrossOmega} \\
\Delta \mathbf{\Omega =}-\left[ \nabla \times \left[ \nabla \times \mathbf{%
\Omega }\right] \right] =\frac{1}{R^{2}}\left( \frac{d^{2}\Omega _{\phi
}(\theta )}{d\theta ^{2}}+\cot \theta \frac{d\Omega _{\phi }(\theta )}{%
d\theta }-(\cot ^{2}\theta -11)\Omega _{\phi }(\theta )\right) \frac{r}{R}%
\mathbf{e}_{\phi }.  \label{Laplacian}
\end{gather}%
From the last relation it follows that it is sufficient to solve equation (%
\ref{eqOm}) up to the first order in $r/R$. We expand the right-hand side of
equation (\ref{eqOm}) in a series in $r/R$:
\begin{gather}
\mathbf{N}\times \lbrack (\mathbf{N}\cdot \vec{\nabla})\mathbf{N}]=\mathbf{E}%
^{lin}(\mathbf{r})\times \left( (\mathbf{E}^{lin}(\mathbf{r})\cdot \nabla )(%
\mathbf{E}^{lin}(\mathbf{r})+\nabla \times \mathbf{\Omega })\right) +O\left(
\frac{r}{R}\right) ^{2}=  \label{fOmega} \\ \nonumber
=\frac{(q_{1}-q_{2})^{2}}{16\pi ^{2}R^{7}}\sin ^{2}\theta \bigg{(} -\frac{%
d^{2}\Omega _{\phi }}{d\theta ^{2}}+4\cot \theta \frac{d\Omega _{\phi }}{%
d\theta }+\\ \nonumber
+(3-6\csc ^{2}\theta )\Omega _{\phi }-\frac{12q_{1}q_{2}}{4\pi
	R(q_{1}-q_{2})\sin \theta }\bigg{)} \frac{r}{R}\mathbf{e}_{\phi }+O\left(
\frac{r}{R}\right) ^{2},   \\
2\frac{1-\xi \left( \gamma N^{2}\right) }{N^{2}}=\frac{8\pi R^{4}}{%
(q_{1}-q_{2})^{2}}\left( 1-\kappa (R)\right) +O\left( \frac{r}{R}\right) .
\label{fOmega1}
\end{gather}%
Thus, we obtain from (\ref{eqOm}) with the use of \ref{Laplacian}), (\ref%
{fOmega}) and (\ref{fOmega1}) a linear differential equation for the
function $\Omega _{\phi }(\theta )$:
\begin{gather}
\left( \kappa (R)-2-\left( \kappa (R)-1\right) \cos 2\theta \right) \frac{%
d^{2}\Omega _{\phi }(\theta )}{d\theta ^{2}}-\left( 4\left( \kappa
(R)-1\right) \sin 2\theta +\cot \theta \right) \frac{d\Omega _{\phi }(\theta
)}{d\theta }+  \notag \\
+\left( 3\kappa (R)\left( \cos 2\theta +3\right) -3\left( \cos 2\theta
+7\right) +\csc ^{2}\theta \right) \Omega _{\phi }(\theta )=\frac{%
24q_{1}q_{2}(\kappa (R)-1)}{4\pi R(q_{2}-q_{1})}\sin \theta ,  \label{eq23}
\end{gather}%
where $\kappa (R)=4\pi \tilde{E}^{(0)}R^{2}/(q_{2}-q_{1})=\xi \left( \gamma
\frac{(q_{2}-q_{1})^{2}}{16\pi ^{2}R^{4}}\right) ,$ and $\xi $ is the
solution (\ref{ksi1}) of Eq. (\ref{ksi}). The general solution of equation (%
\ref{eq23}) in the class of functions regular in $\theta $ has the form:
\begin{equation}
\Omega _{\phi }(\theta )=\frac{12q_{1}q_{2}}{4\pi (q_{1}-q_{2})R}\frac{%
\kappa (R)-1}{3\kappa (R)-2}\sin \theta \cos ^{2}\theta +C_{1}\sin \theta
(1+7\cos (2\theta )+\frac{4}{\kappa (R)}\sin ^{2}\theta ).  \label{45}
\end{equation}%
The term with the constant $C_{1}$ satisfies the homogeneous equation
obtained from (\ref{eq23}) by omitting its right-hand side. Consequently,
this solution determines a field that is not generated by the source, and
therefore we discard it. The condition $C_{1}=0$\ can be represented also in
the form $\Omega _{\phi }\left(\pi /2 \right) =0$. By
substituting (\ref{45}) with $C_{1}=0$ into (\ref{nablacrossOmega}) we have
\begin{equation}\nonumber
\left[ \nabla \times \mathbf{\Omega }\right] =\frac{24q_{1}q_{2}}{4\pi
(q_{2}-q_{1})R^{2}}\frac{\kappa (R)-1}{2\kappa (R)-3}\frac{(\mathbf{r}\cdot
\mathbf{R})}{R^{2}}\left( 2\frac{(\mathbf{r}\cdot \mathbf{R})}{R^{2}}\frac{%
\mathbf{R}}{R}-\frac{\mathbf{r}}{R}\right) +O\left( \frac{r}{R}\right) ^{3}.
%\label{curlOmega}
\end{equation}%
Then
\begin{gather}\nonumber
\mathbf{N}=\mathbf{E}^{lin}(\mathbf{r})+\nabla \times \mathbf{\Omega }=%
\mathbf{N}^{(0)}+\mathbf{N}^{(1)}+\mathbf{N}^{(2)}+O\left( \frac{r}{R}%
\right) ^{3}, \\ \nonumber
\mathbf{N}^{(0)}=\frac{q_{2}-q_{1}}{4\pi R^{2}}\frac{\mathbf{R}}{R},\quad
\mathbf{N}^{(1)}=\frac{q_{1}+q_{2}}{4\pi R^{2}}\left( \frac{\mathbf{r}}{R}-3%
\frac{(\mathbf{r},\mathbf{R})}{R^{2}}\frac{\mathbf{R}}{R}\right) ,  \notag \\ \nonumber
\mathbf{N}^{(2)}=\frac{3(q_{1}-q_{2})}{8\pi R^{2}}\left[ 2\alpha (R)\frac{(%
\mathbf{r},\mathbf{R})}{R^{2}}\frac{\mathbf{r}}{R}+\left( \frac{r}{R}\right)
^{2}\frac{\mathbf{R}}{R}-5\beta (R)\frac{(\mathbf{r},\mathbf{R})^{2}}{R^{4}}%
\frac{\mathbf{R}}{R}\right] .  \notag
\end{gather}%
where
\begin{equation}\nonumber
\alpha (R)=\frac{(q_{1}+q_{2})^{2}+\frac{4q_{1}q_{2}}{2\kappa (R)-3}}{%
(q_{1}-q_{2})^{2}},\quad \beta (R)=\frac{5(q_{1}^{2}+q_{2}^{2})+2q_{1}q_{2}%
\left( 3+\frac{8}{2\kappa (R)-3}\right) }{5(q_{1}-q_{2})^{2}}.
\end{equation}%
For $N^{2}$ we have
\begin{gather}\nonumber
N^{2}=N_{20}+N_{21}+N_{22}+O\left( \frac{r}{R}\right) ^{3}, \\ \nonumber
N_{20}=\frac{(q_{1}-q_{2})^{2}}{16\pi ^{2}R^{4}},\quad N_{21}=\frac{%
q_{1}^{2}-q_{2}^{2}}{4\pi ^{2}R^{4}}\frac{(\mathbf{r},\mathbf{R})}{R^{2}},
\notag \\ \nonumber
N_{22}=\frac{3}{4\pi ^{2}R^{4}}\frac{(\mathbf{r},\mathbf{R})^{2}}{R^{4}}%
\left( q_{1}^{2}+q_{2}^{2}+q_{1}q_{2}\frac{2\kappa (R)-1}{2\kappa (R)-3}%
\right) -\frac{1}{8\pi ^{2}R^{4}}(q_{1}^{2}+q_{2}^{2}-4q_{1}q_{2})\left(
\frac{r}{R}\right) ^{2}.  \notag
\end{gather}%
We expand expression (\ref{vExi}) in series within the order $(r/R)^2$:
\begin{equation}\nonumber
\mathbf{E}=\mathbf{N}\xi \left( \gamma N^{2}\right) =\mathbf{\tilde{E}}%
^{(0)}+\mathbf{\tilde{E}}^{(1)}+\mathbf{\tilde{E}}^{(2)}+O\left( \frac{r}{R}%
\right) ^{3}.
\end{equation}%
The first two terms $\mathbf{\tilde{E}}^{(0)}$ and $\mathbf{\tilde{E}}^{(1)}$
are determined by the (\ref{result}). For second-order correction $\mathbf{%
\tilde{E}}^{(2)}$ we obtain
\begin{gather}\nonumber
\mathbf{\tilde{E}}^{(2)}=\xi \left( \gamma N_{20}\right) \mathbf{N}%
^{(2)}+\gamma N_{21}\xi ^{\prime }\left( \gamma N_{20}\right) \mathbf{N}%
^{(1)}+\left( \gamma \xi ^{\prime }\left( \gamma N_{20}\right) N_{22}+\frac{\gamma
}{2}\xi ^{\prime \prime }\left( \gamma N_{20}\right) N_{21}^{2}\right)
\mathbf{N}^{(0)}.  \notag
\end{gather}%
Differentiating (\ref{ksi}) we obtain expressions
\begin{equation}\nonumber
\xi ^{\prime }(x)=-\frac{\xi ^{3}(x)}{2+3x\xi (x)^{2}},\quad \xi ^{\prime
\prime }(x)=12\frac{(1+x\xi ^{2}(x))\xi ^{5}(x)}{(2+3x\xi ^{2}(x))^{3}}.
\end{equation}%

This finally results in the second-power correction to (\ref{result}):
\begin{equation}
\widetilde{\mathbf{E}}^{(2)}=\frac{3}{8\pi R^{2}}\left( c(R)\left[ 2\frac{(%
\mathbf{r}\cdot \mathbf{R})}{R^{2}}\frac{\mathbf{r}}{R}+\frac{\mathbf{R}}{R}%
\left( \frac{r}{R}\right) ^{2}\right] -5d(R)\frac{(\mathbf{r}\cdot \mathbf{R}%
)^{2}}{R^{4}}\frac{\mathbf{R}}{R}\right),  \label{cd}
\end{equation}%
where
\begin{gather}\nonumber
c(R)=\frac{1}{3\kappa (R)}\left( \frac{(q_{1}+q_{2})^{2}}{q_{1}-q_{2}}+\frac{%
q_{1}^{2}-4q_{1}q_{2}+q_{2}^{2}}{q_{1}-q_{2}}\frac{2\kappa (R)}{3\kappa (R)-2%
}\right) , \\ \nonumber
d(R)=\frac{\kappa (R)}{5(q_{1}-q_{2})(2\kappa (R)-3)}\bigg{(}%
8(q_{1}+q_{2})^{2}\kappa
^{3}(R)-4(13(q_{1}^{2}+q_{2}^{2})+14q_{1}q_{2})\kappa ^{2}(R)+  \notag \\ \nonumber
+2(47(q_{1}^{2}+q_{2}^{2})+14q_{1}q_{2})\kappa
(R)-11(5(q_{1}^{2}+q_{2}^{2})-2q_{1}q_{2})\bigg{)}.
\end{gather}%
Note that (\ref{cd}) obeys the first Maxwell equation $\left[ \nabla \times
\widetilde{\mathbf{E}}^{(2)}\right] =0$ identically for any coefficients $c$
and $d$. 

Similarly to the coefficient $(q_{1}-q_{2})$ in the third (quadrupole) term
in (\ref{lin-quadrupole}), the coefficients $c(R)$ and $d(R)$ are odd under
the permutation $q_{1}\leftrightarrow q_{2}.$ Note that the seeming
singularity at $q_{1}=q_{2}$ cancels from these coefficients due to the
equality $\kappa (R)=1$\ that holds in this case. In the linear limit $%
\gamma =0$ one also has $\kappa (R)=1,$ and $c(R)$ and $d(R)$\ turn both
into $q_{1}-q_{2},$ so that (\ref{cd}) turns into the last (quadrupole)
term in the expansion (\ref{lin-quadrupole}) of the linear field.

\section{Concluding remarks}

We were working within the simplest nonlinear electrodynamics with the
self-interaction of the fourth power of electromagnetic field (\ref{quartic}%
), which, if needed, may be thought of as resulting from the first
nontrivial term of expansion of the Euler-Heisenberg effective Lagrangian in
powers of its background field argument $\mathfrak{F}$, while the other
field invariant is kept vanishing $\mathfrak{G=}0.$ In this case\ the
coefficient $\gamma ,$ whose dimensionality is [\textit{length}$^{4}],$ that
determines the strength of nonlinearity is expressed as (\ref{gamma}) in
terms of the electron mass and charge. Otherwise it may be considered as
arbitrary. Anyway, in our calculation a smallness of $\gamma ^{1/4}$ as
compared to the two other quantities $r$ and $R$\ carrying the
dimensionality of length was nowhere assumed.

We considered the electrostatic problem of interaction between two
point-like charges $q_{1}$ and $q_{2}$\ placed in the points $\mathbf{r}=\pm
$ $\mathbf{R}$ by solving the nonlinear Maxwell equation (\ref{two-body}),
which follows from the least action principle for the Lagrangian (\ref{L}),
together with the standard Bianchi identity (\ref{rot}). For the small
separation between the charges, $R\ll r,$ we found the electric field (\ref%
{result1}) in the approximation, linear with respect to the ratio $R/r$, which serves the nonlinear extension of the usual dipole field. The
result for the corresponding scalar potential is Eq. (\ref{pot1}). The
lines-of-force and equipotential-curves pattern is shown in Fig. 1 in the
configuration space $\mathbf{r}$ with the parameters chosen in such a way as
to make the nonlinearity effect best pronounced. For large separation
between the charges, $R\gg r,$ we found the electric field in the
approximation, linear (\ref{result}) with respect to the ratio $r/R$, and quadratic (\ref{cd}).

Using the two opposite reprentations (\ref{result1}) and (\ref{result}) we
can get a rough estimate for the behaviour of the field-energy (\ref{P0}), (%
\ref{theta}) in the asymptotic regime $R\rightarrow 0,$ where the two point
charges approach each other infinitely close. According to that estimate, in
this regime the energy of the system of two point charges can be presented
as
\begin{equation}
P^{0}=a+b R^{\frac{1}{3}},  \label{P0As}
\end{equation}
where $a$ and $b$ are finite constants depending only on the charges $q_{1}$
and $q_{2},$ and on the self-coupling constant $\gamma $. The $R$%
-independent term $a$ is the self-energy of the united point-like charge
with the value $q_{1}+$ $q_{2}.$ This is finite, as established in \cite%
{CosGitSha2013a}. The behaviour (\ref{P0As}) is rigorously confirmed
following a quite different procedure to be published elsewhere. Although
the energy is finite in the limit $R=0,$ the force $\mathbf{F}$ between the
two charges defined as the derivative of the energy with respect to the
distance is weakly infinite:%
\begin{equation*}
\mathbf{F=}\frac{dP^{0}}{d\mathbf{R}}=\frac{\mathbf{R}}{R^{\frac{5}{3}}}%
\frac{b}{3}.
\end{equation*}%
This formula replaces, in the given nonlinear model, the Coulomb law $%
\mathbf{F}\sim \mathbf{R}\cdot R^{-3}$ for the force between two point
charges. The power $2/3$ here is determined by the power $2$ in the
self-interaction in (\ref{quartic}).

\section*{Acknowledgements}

Supported by RFBR under Project 17-02-000317, and by the TSU Competitiveness
Improvement Program, by a grant from \textquotedblleft The Tomsk State
University D. I. Mendeleev Foundation Program\textquotedblright .

%% The bibliography section

\end{document}